\documentclass{appolb}
\usepackage{graphicx}
\usepackage{amssymb}
\usepackage[utf8]{inputenc}
\usepackage{comment}
\begin{document}
\title{PROSA PDFs and astrophysical applications%
\thanks{Presented at ``Diffraction and Low-x 2018'', Reggio Calabria, Italy, 27 August - 1~September 2018}%
}
\author{Maria Vittoria Garzelli for the PROSA collaboration\thanks{S.-O.~Moch, O.~Zenaiev, A.~Cooper-Sarkar, A.~Geiser, K.~Lipka, G.~Sigl} 
\address{Institute for Theoretical Physics, University of Tuebingen, Auf~der~Morgenstelle~14, D-72076 Tuebingen, Germany}
%\\
%{Third Author of different affiliation
%}
%the Name(s) of other Author(s)
%\address{affiliation}
}
\maketitle
\begin{abstract}
The PROSA parton distribution function fit was the first one appeared
in the literature incorporating data on open charm and open bottom
production at LHCb, in order to reduce the uncertainties on gluons
and sea quarks at low $x$'s ($x$ $<$ $10^{-4}$).
We discuss aspects of the PROSA PDFs of particular
relevance for their usage in the field of Neutrino Astronomy, and their
application in the computation of prompt neutrino fluxes. 
\end{abstract}
%\PACS{PACS numbers come here}
  
\section{Introduction}
Constraining Parton Distribution Functions (PDF) at low values of the longitudinal momentum fraction $x$ is crucial for a series of high-energy applications, ranging from the development of new colliders with increasing center-of-mass energy $\sqrt{s}$, to the interpretation of data from high-energy astroparticle physics. 

At the core of all present PDF fits there are the Deep-Inelastic-Scattering (DIS) data collected at the HERA $e p$ collider, which allow to probe $x$ values in the range $10^{-4}$ $\lesssim$ $x$ $\lesssim$ 0.1 . Some of these data, i.e. those on the longitudinal structure functions $F_L$, allow to extend this range to $x$~$\gtrsim$~$4~\cdot~10^{-5}$, although with big uncertainties. This $x$ coverage is enough for many analyses at the Large Hadron Collider (LHC), with the ATLAS and CMS detectors mostly focused 
on particle production at central rapidity $|y|$~$<$~2.5 and large transverse momentum $p_T$. However, interpreting experimental information on particle production at larger $|y|$ / larger $\sqrt{s}$ requires the development of PDF fits reliable even below $x$ $\sim$ $10^{-5}$. In fact, the higher is the $\sqrt{s}$ of a $pp$ collision, the lower are the $x$ values which can characterize the partons involved in the elementary scattering processes inherent to it. Additionally, at  fixed $\sqrt{s}$, particles at larger $|y|$ are, on average, produced by scattering of partons with more extreme $x$'s, i.e. a very small $x$ for the parton from one hadron in combination with a very large $x$ for the parton from the other hadron. 

The need of investigating these kinematical regions is particularly evident in cosmic ray (CR) physics applications. First of all, the most energetic CRs reaching the Earth have energies well above those reachable at present-day accelerators: the CR spectrum extends up to laboratory energies  $E_{lab} \sim 10^{11}$ GeV, whereas the present LHC $pp$ $\sqrt{s}$ corresponds to $E_{lab}~\sim$~$10^8$~GeV. Furthermore, as follows from geometry considerations, most CR interactions typically involve small momentum transfers $Q^2$ and lead to particle production in the forward region, which corresponds to more extreme $x$ values than in case of central production at large $Q^2$. 

The question of PDF behaviour at low $x$'s is still very debated, and the number of open issues further arises when considering the combination (low~$x$, low $Q^2$). 
%In this contribution 
We describe the investigations in this respect by the PROSA collaboration, which, first among the various PDF collaborations, proposed the idea of using the data on open heavy-meson hadroproduction recorded by the LHCb experiment, which span rapidities in the 2 $< y <$ 4.5 interval, divided in five sub-intervals of equal size, to constrain PDFs in the $x \in$ [$10^{-6}$, $10^{-4}$]  range~\cite{Zenaiev:2015rfa}. 

\section{The PROSA PDF fit}
The PROSA PDF fit employs the combined set of neutral current (NC) and charded current (CC) inclusive  DIS data collected by the H1 and ZEUS Collaborations at HERA at $\sqrt{s}$ = 320 GeV, which was also the basis of the HERAPDF1.0 PDF fit~\cite{Aaron:2009aa}. These data are directly sensitive to the valence and sea quark distributions and allow to put constraints on the gluon distribution through scaling violations (down to $x \gtrsim 10^{-3}$). The fit also includes combined H1 and ZEUS semi-inclusive data on 
charm production in NC DIS~\cite{Abramowicz:1900rp}, plus ZEUS data on bottom production in DIS~\cite{Abramowicz:2014zub}. These data allow to further constrain gluon and sea distributions (down to $x \gtrsim 10^{-4}$), as well as the values of the charm and bottom masses. Additionally, the PROSA fit includes the LHCb data on open $D^{\pm}$, $D^0$, $\bar{D}^0$, $D_s^{\pm}$, $\Lambda_c^+$, $B^+$, $B^0$, $B_s^0$ meson production at $\sqrt{s}$~=~7~TeV~\cite{Aaij:2013mga, Aaij:2013noa}. The LHCb data allow to constrain gluon and sea distributions in the $x \in$ [$5 \cdot 10^{-6}$, $10^{-3}$] and [$10^{-2}$,~$1$] ranges. 

\begin{figure}[htb]
\begin{center}
\includegraphics[width=0.325\textwidth]{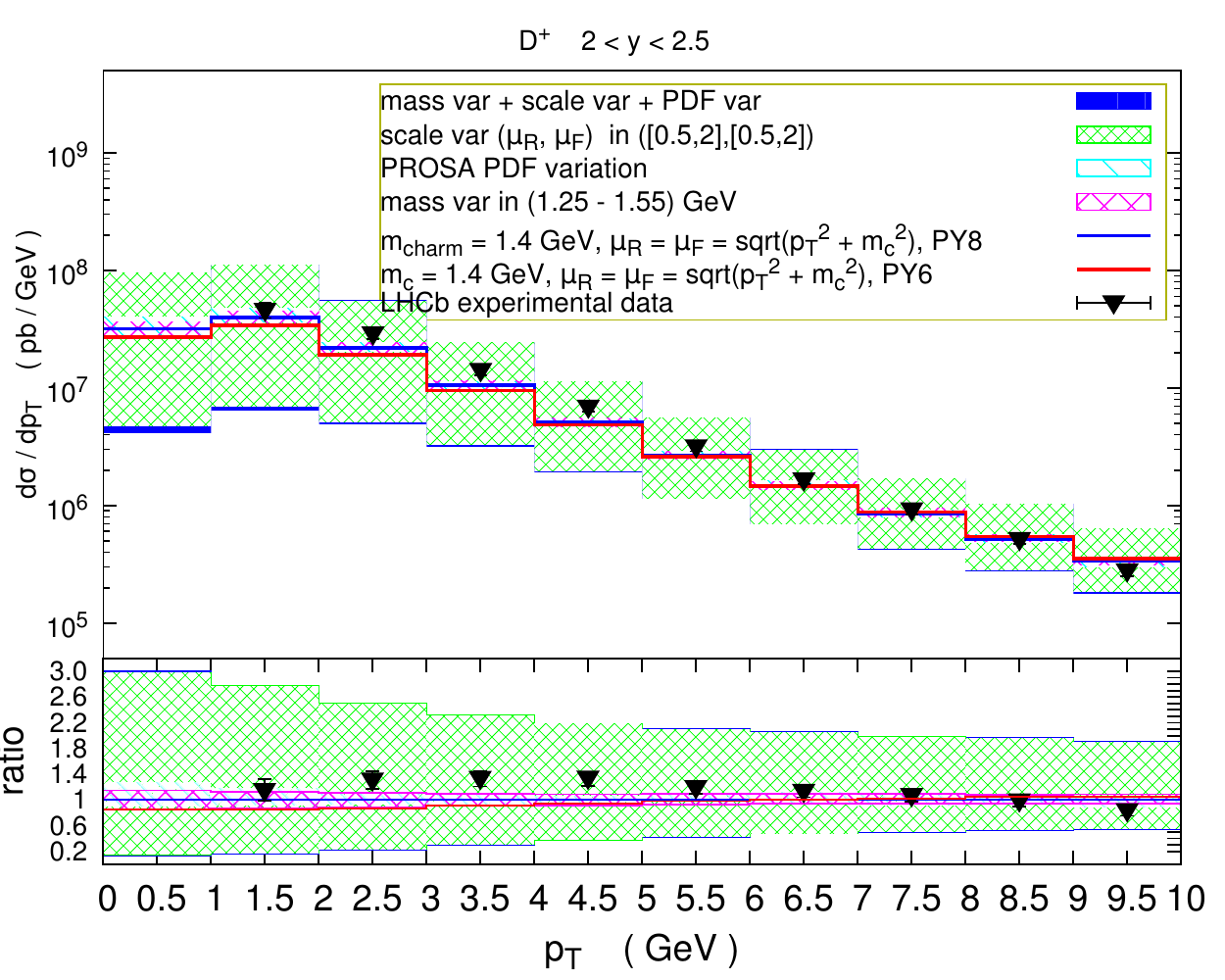}
\includegraphics[width=0.325\textwidth]{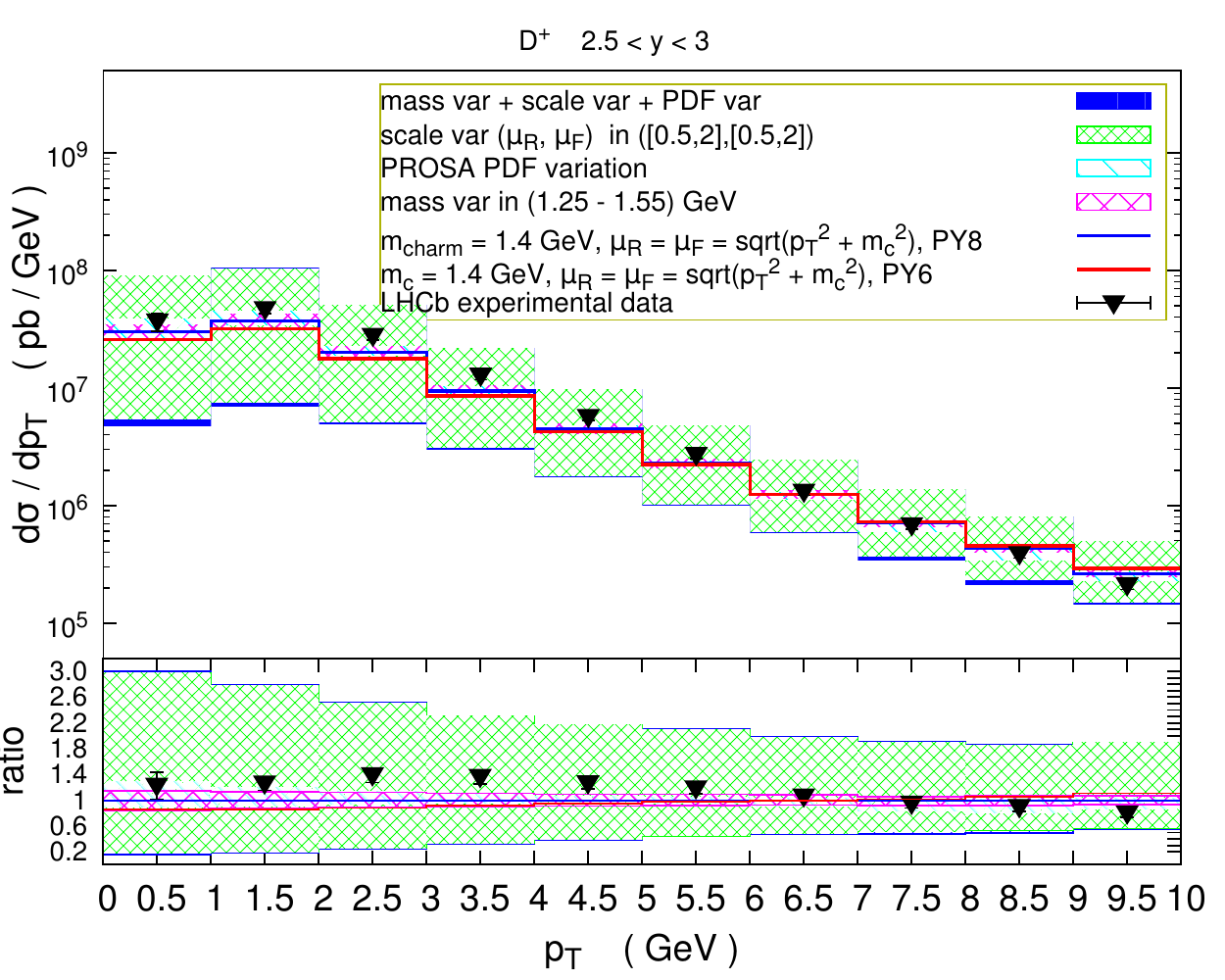}
\includegraphics[width=0.325\textwidth]{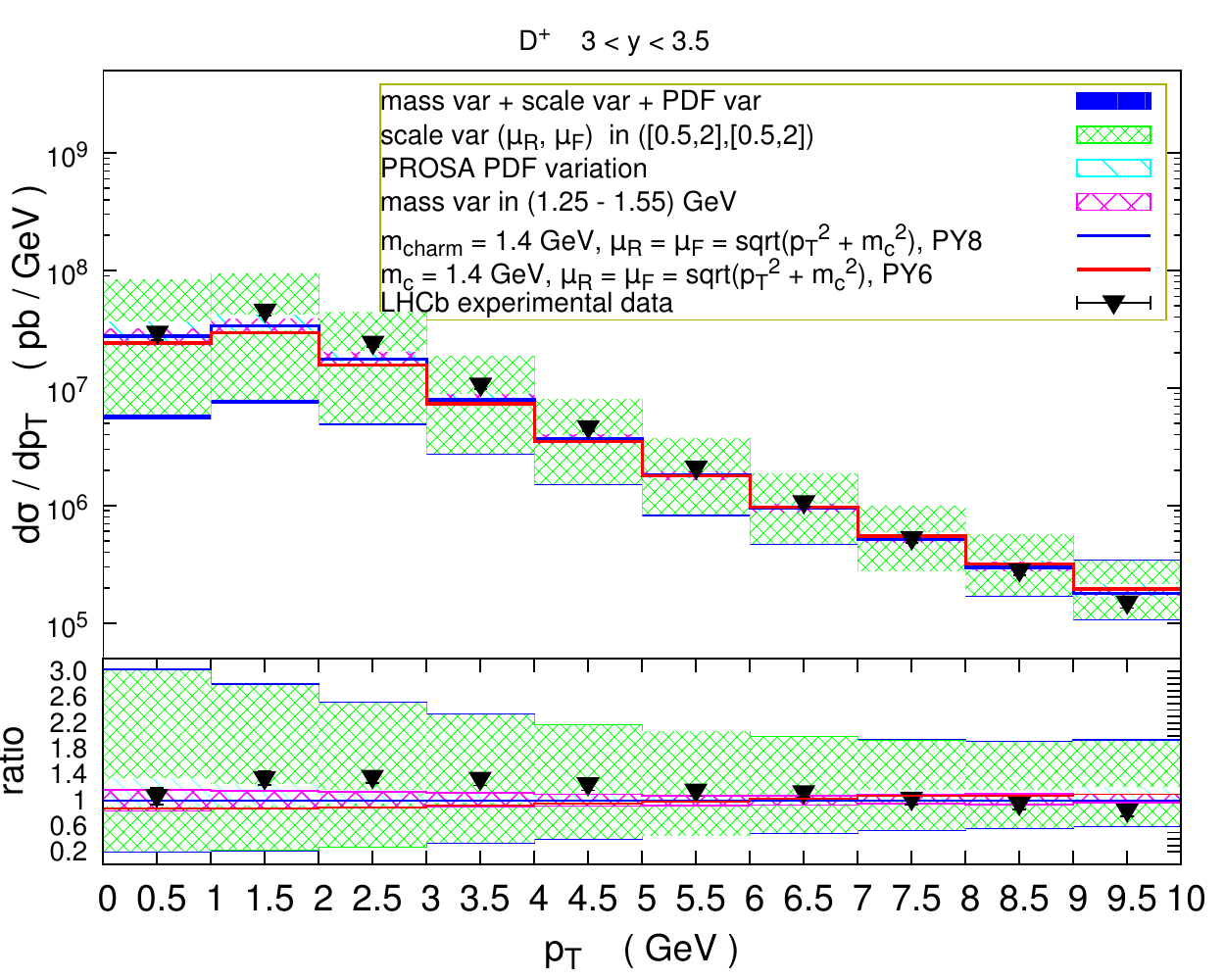}\\
\includegraphics[width=0.325\textwidth]{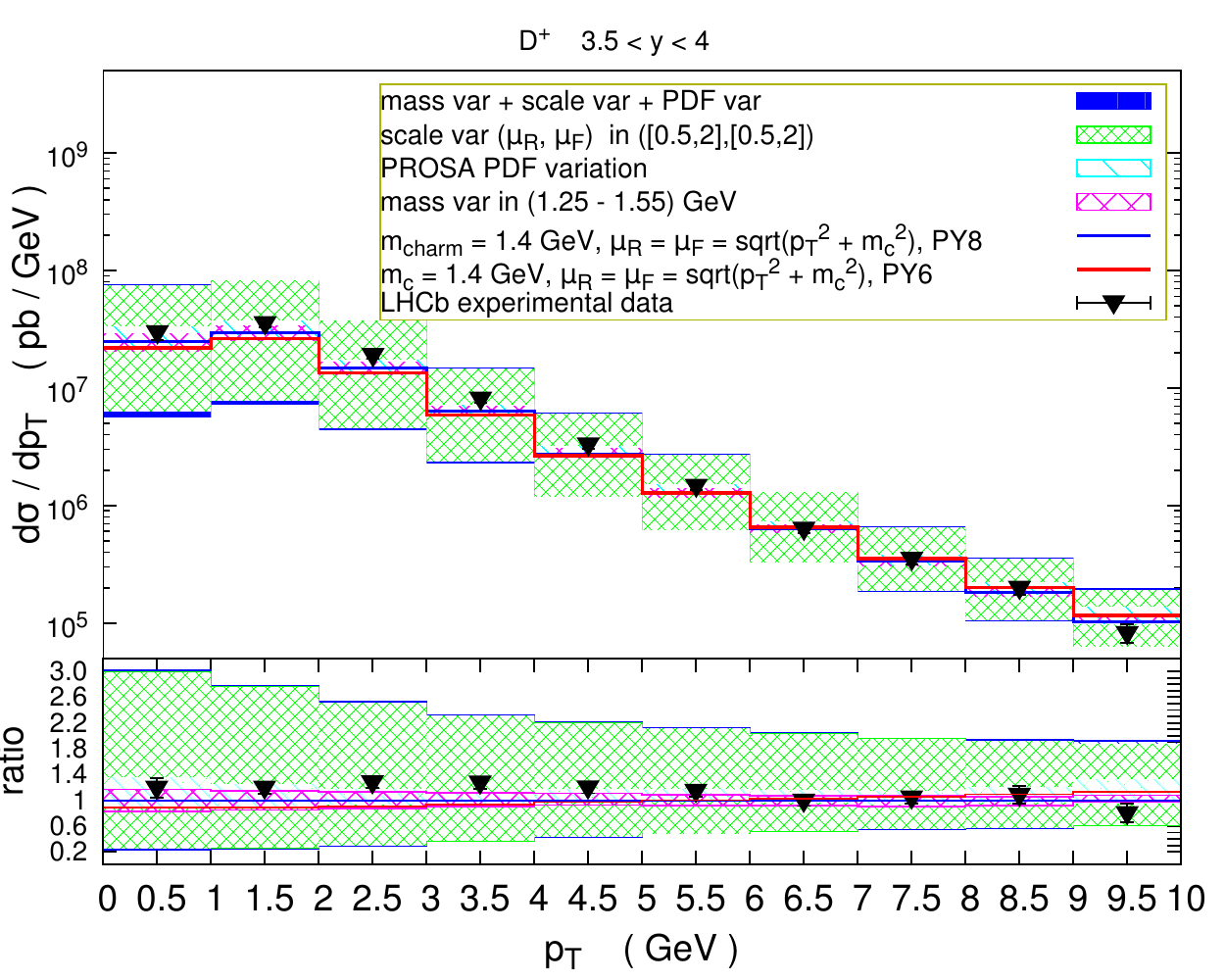}
\includegraphics[width=0.325\textwidth]{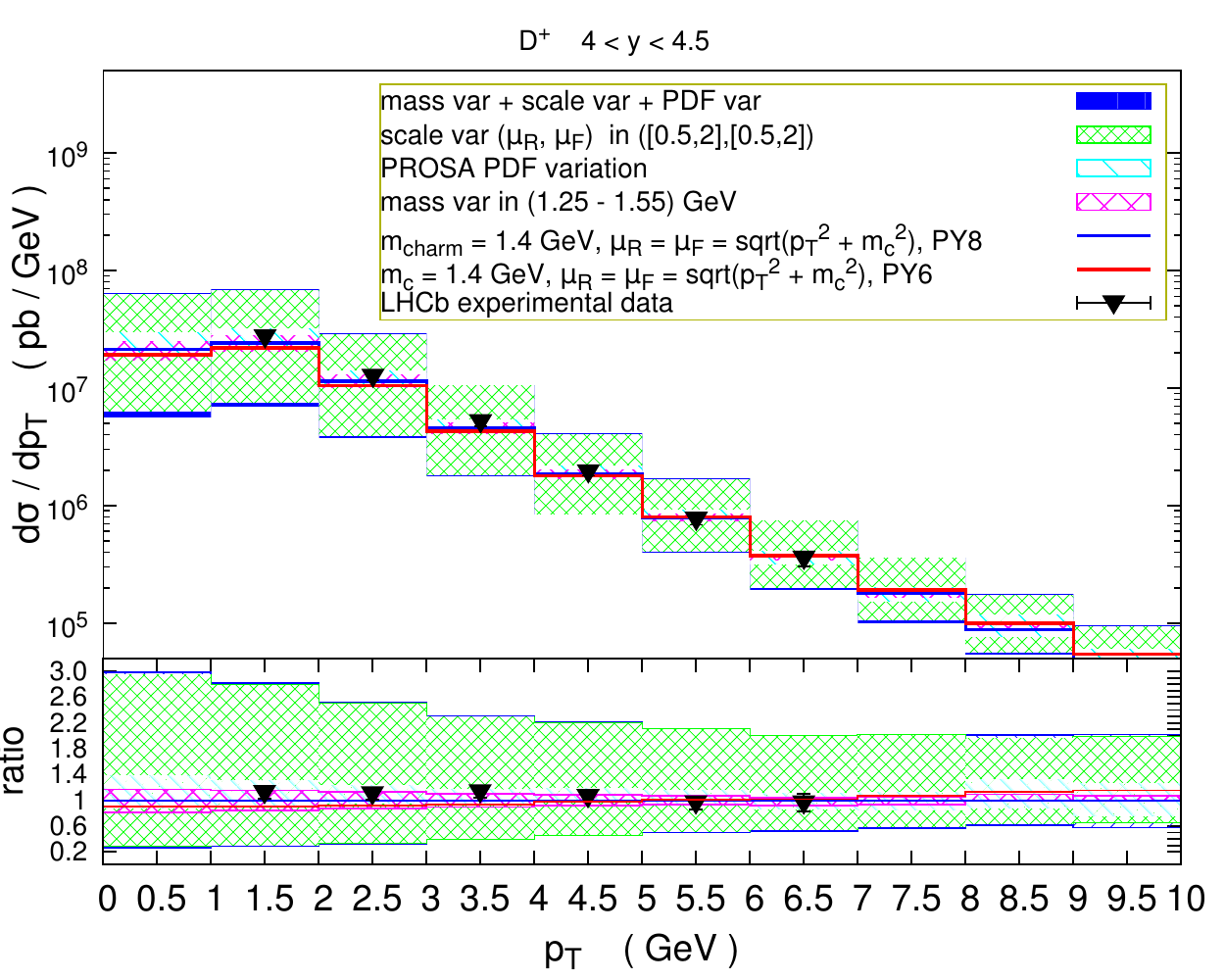}
\vspace{-0.5cm}
\end{center}
\caption{PROSA d$\sigma$/d$p_T$ predictions for $pp$ $\rightarrow D^\pm$ + X at $\sqrt{s}$ = 5 TeV vs. LHCb experimental data. Each panel refers to a different $y$ sub-interval. Theoretical uncertainty bands refer to scale, PDF and charm mass variation. In the lower panels, ratios with respect to the central theoretical predictions are reported.} 
\label{fig:lhcb5}
\end{figure}

Two variants of the fit are proposed: in the first one the absolute values of the differential cross-sections $d \sigma / d p_T$, measured in the five available LHCb $y$ sub-intervals, are used. In the second variant, for each $p_T$ bin, ratios of cross-sections in different $y$ sub-intervals are employed, considering as reference sub-interval the central LHCb one, with 3~$<$~$y$~$<$~3.5. The use of these ratios allows to shrink the uncertainty band related to renormalization and factorization scale  ($\mu_R$ and $\mu_F$) variations, which are the dominant uncertainties in the next-to-leading-order (NLO) calculations of open charm and bottom production at LHCb.
As a consequence, the PDF uncertainty bands are smaller in the second variant of the fit than in the first one. However, they have a large overlapping region. Therefore the two variants of the fit can be considered, overall, consistent one with each other. Thus, in the following, we limit ourselves to present theoretical predictions obtained by employing as input the second variant of the PROSA PDF fit.

\begin{figure}
\begin{center}
\includegraphics[width=0.325\textwidth]{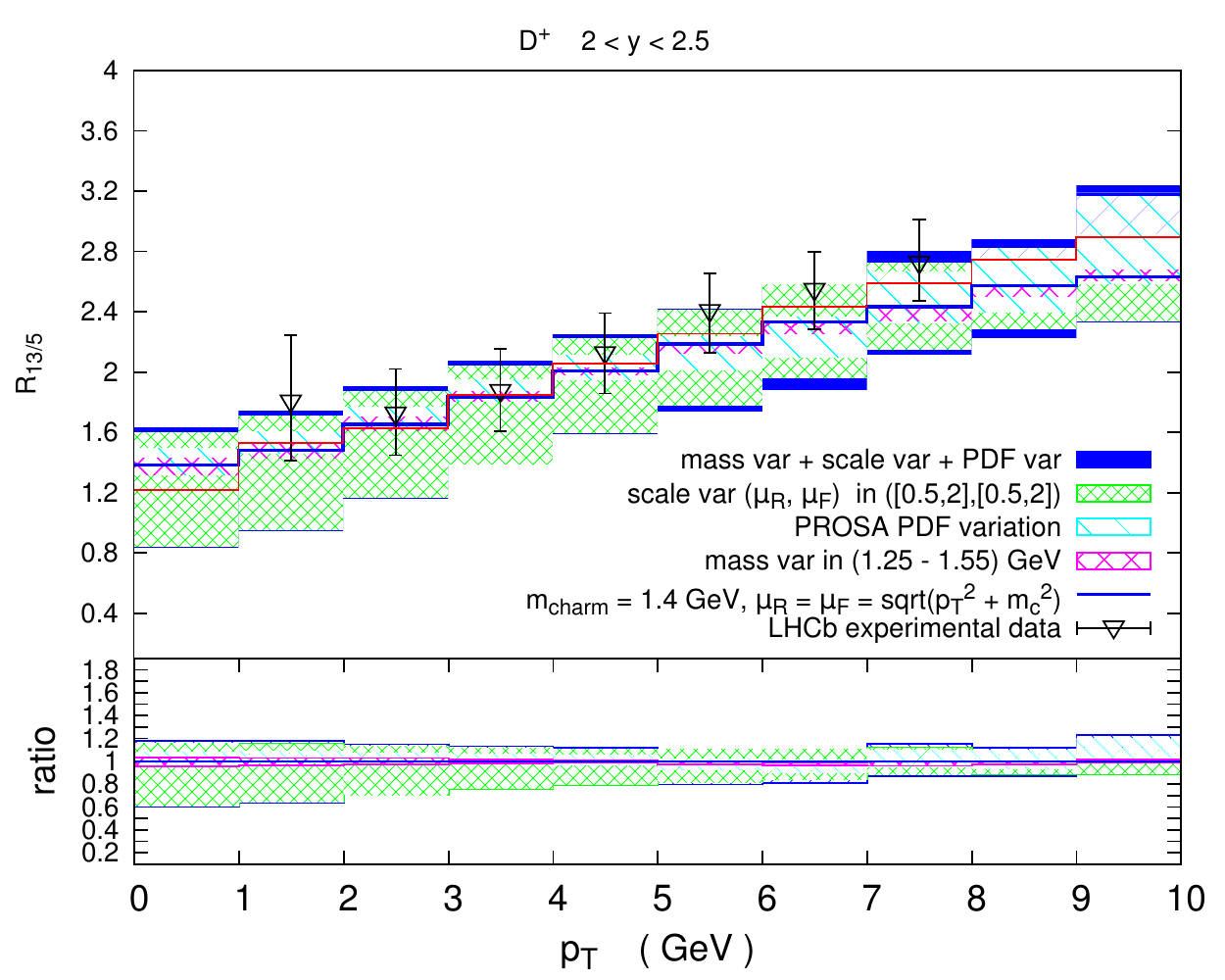}
\includegraphics[width=0.325\textwidth]{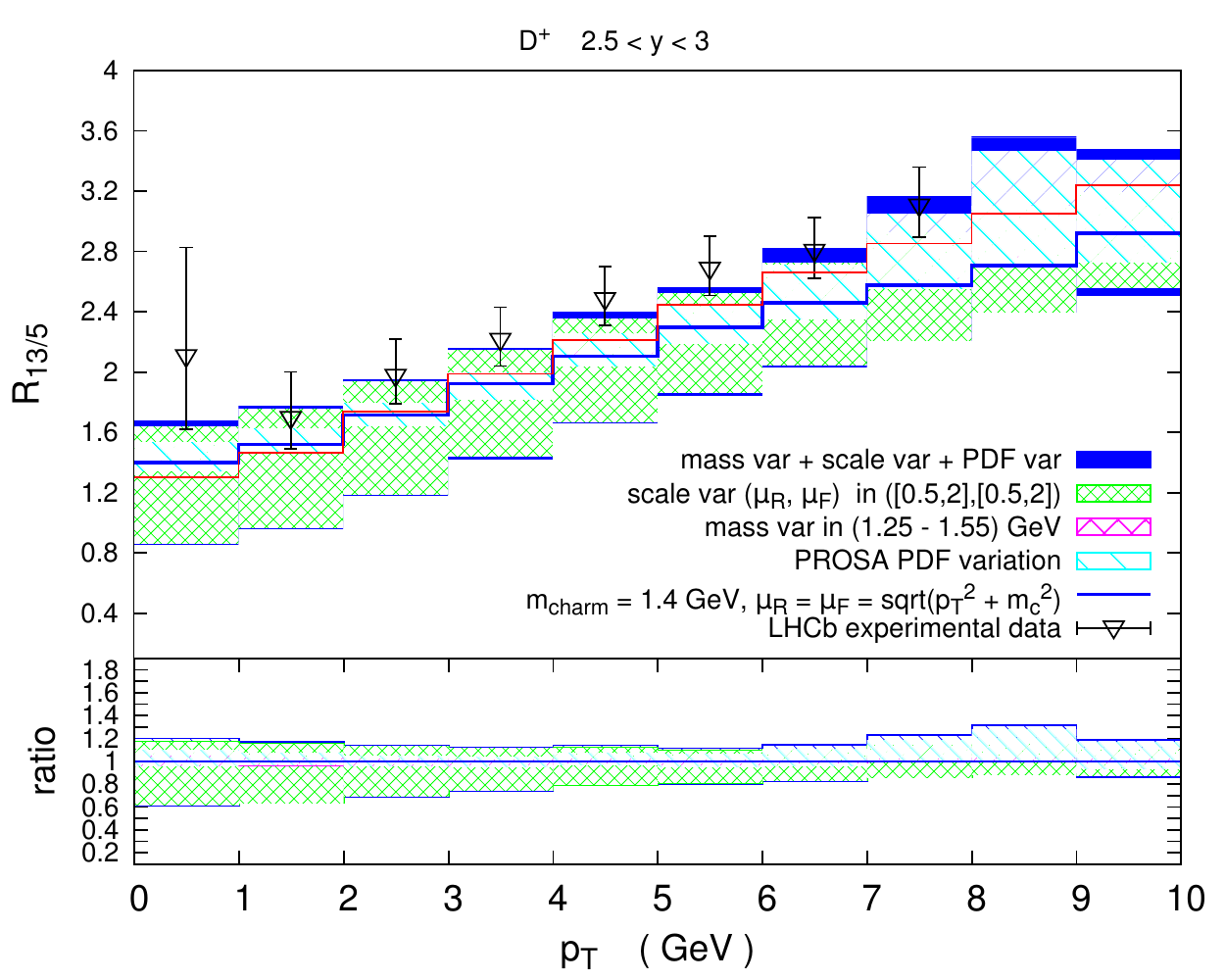}
\includegraphics[width=0.325\textwidth]{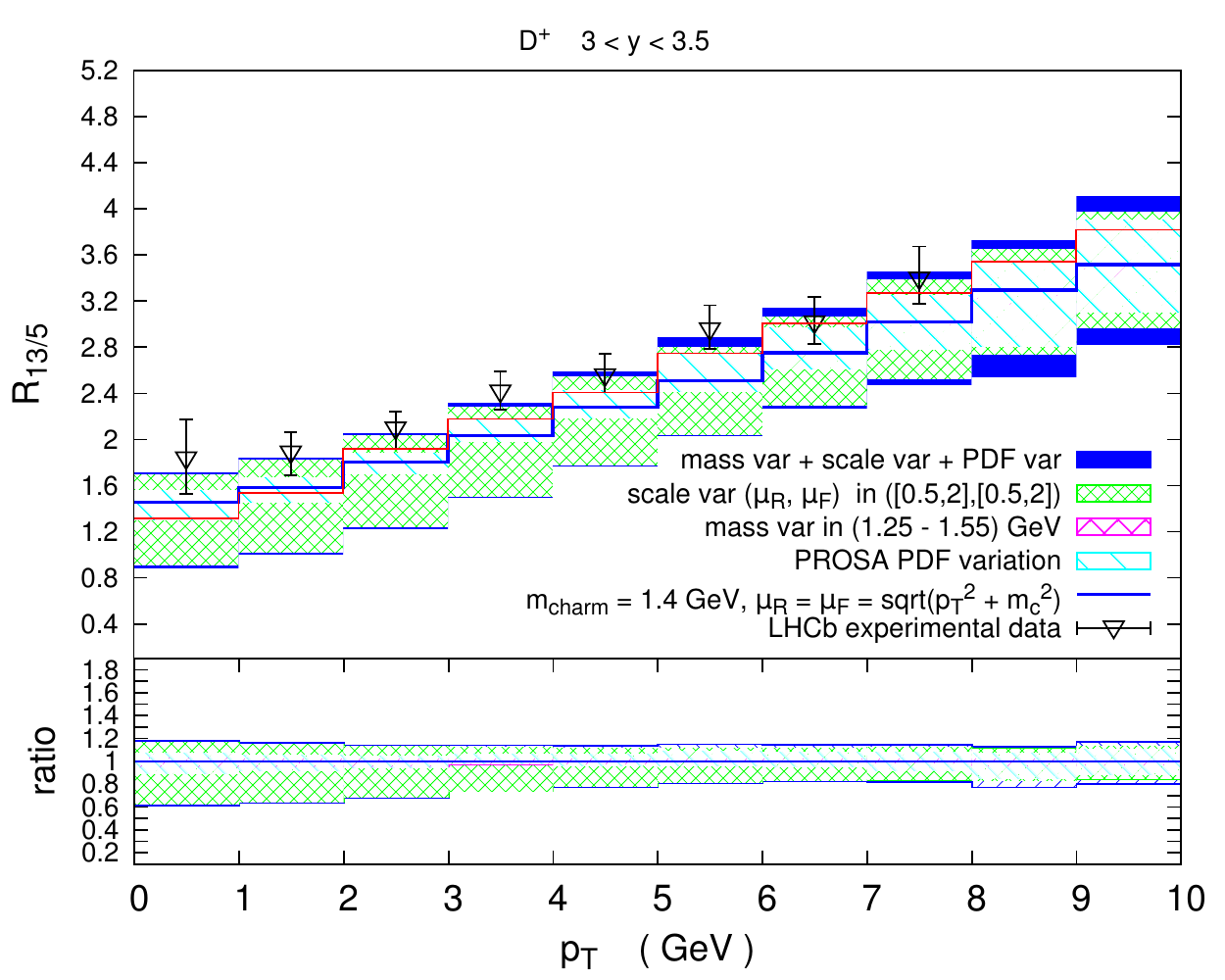}\\
\includegraphics[width=0.325\textwidth]{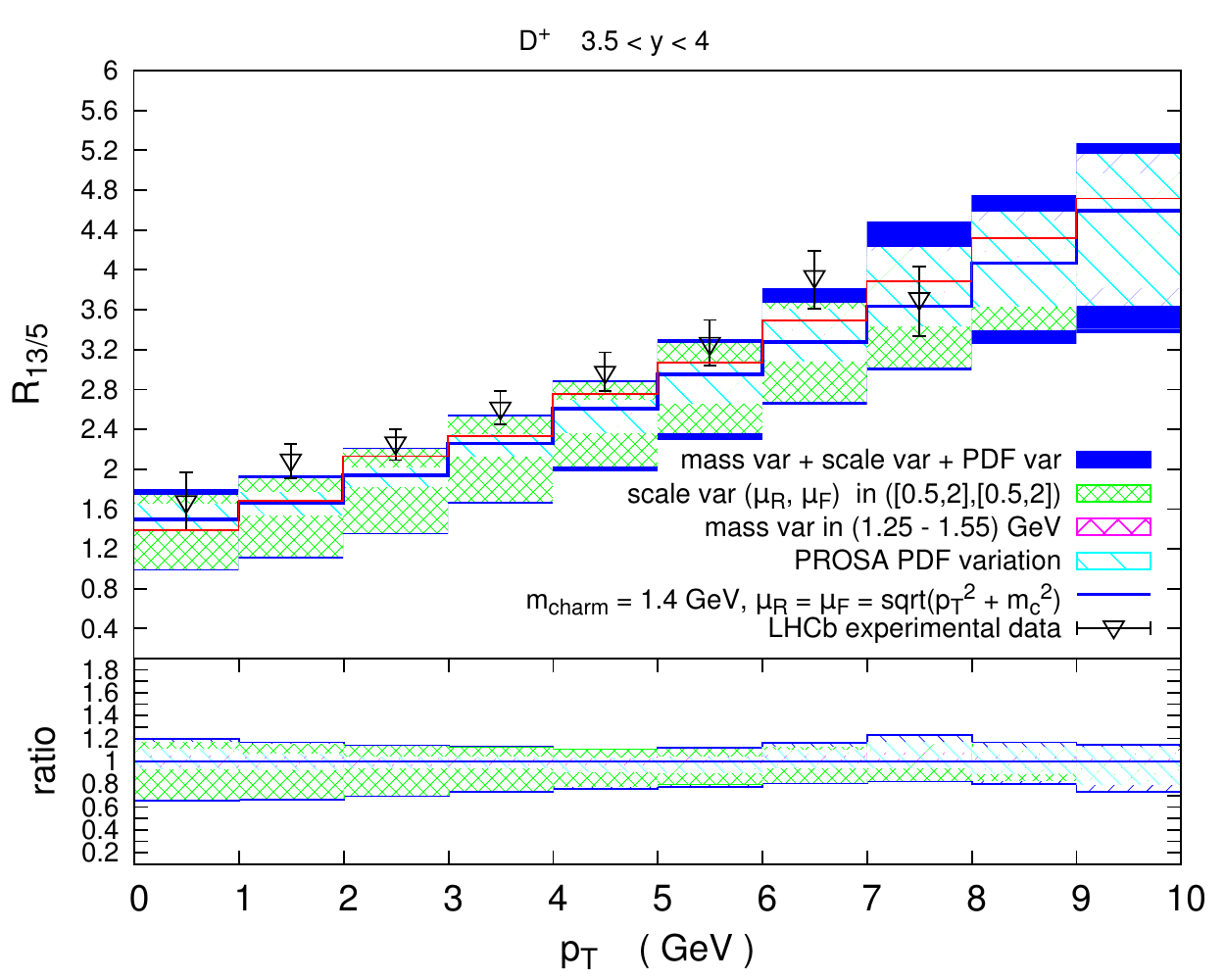}
\includegraphics[width=0.325\textwidth]{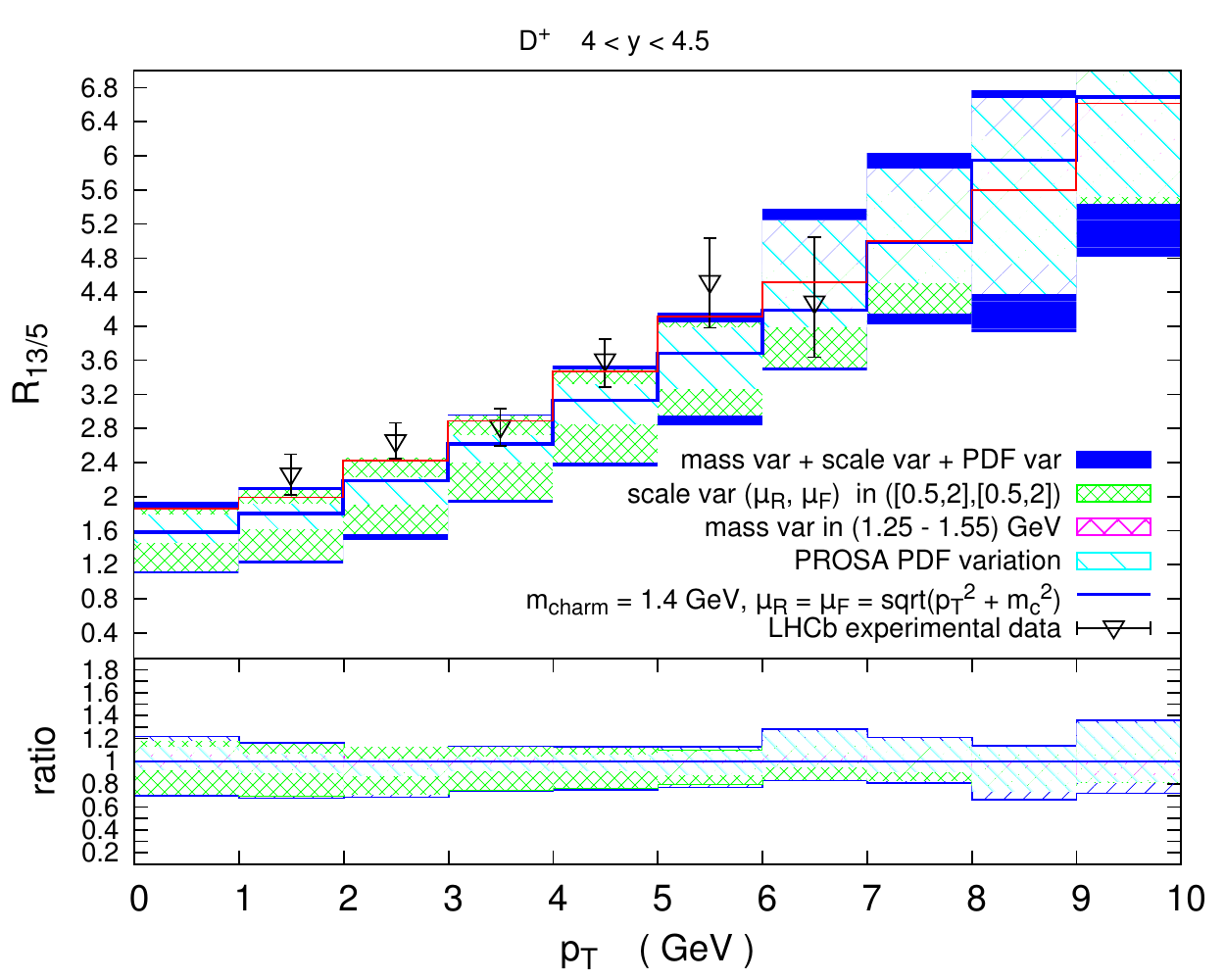}
\vspace{-0.5cm}
\end{center}
\caption{Ratio of PROSA d$\sigma$/d$p_T$ predictions for $pp$ $\rightarrow D^\pm$ + X at $\sqrt{s}$ = 13 TeV and 5 TeV vs. LHCb experimental data. Each panel refers to a different $y$ sub-interval. Theoretical uncertainty bands are as in Fig.~1. Predictions from an independent GM-VFNS calculation, using CT14nlo PDFs, are also reported (red lines). }
\label{fig:lhcb513}
\end{figure}

We test the fit validity 
by comparing our theoretical predictions to further experimental data, not included in the fit itself. Our predictions employing the PROSA PDFs, are based on a perturbative NLO QCD computation of the hard-scattering matrix-elements matched, according to the POWHEG method~\cite{Nason:2004rx, Frixione:2007nw}, to the parton shower and hadronization algorithms provided by the {\texttt{PYTHIA8}} event generator~\cite{Sjostrand:2007gs}. 
In Fig.~\ref{fig:lhcb5} we show their comparison to LHCb experimental data at $\sqrt{s}$ = 5 TeV~\cite{Aaij:2016jht}. Furthermore, in Fig.~\ref{fig:lhcb513} we show ratios of predictions at $\sqrt{s}$~=~13 and 5 TeV compared with the LHCb experimental data from the same paper. Here, besides our theoretical predictions with the PROSA PDFs, we include those of a completely indipendent computation in the General-Mass Variable-Flavour-Number-Scheme (GM-VFNS)~\cite{Benzke:2017yjn}, employing the CT14nlo VFNS central PDF set~\cite{Dulat:2015mca} and the {\texttt{KKK08}} Fragmentation Functions~\cite{Kneesch:2007ey}. The latter predictions turn out to always lie within the uncertainty bands of the former. The experimental data lie in the upper side of the uncertainty band of the PROSA theoretical predictions. Considering their uncertainties, they are compatible with the theory predictions in all bins.

\section{Applications to High-Energy Astroparticle Physics}

\begin{figure}
\begin{center}
\includegraphics[width=0.48\textwidth]{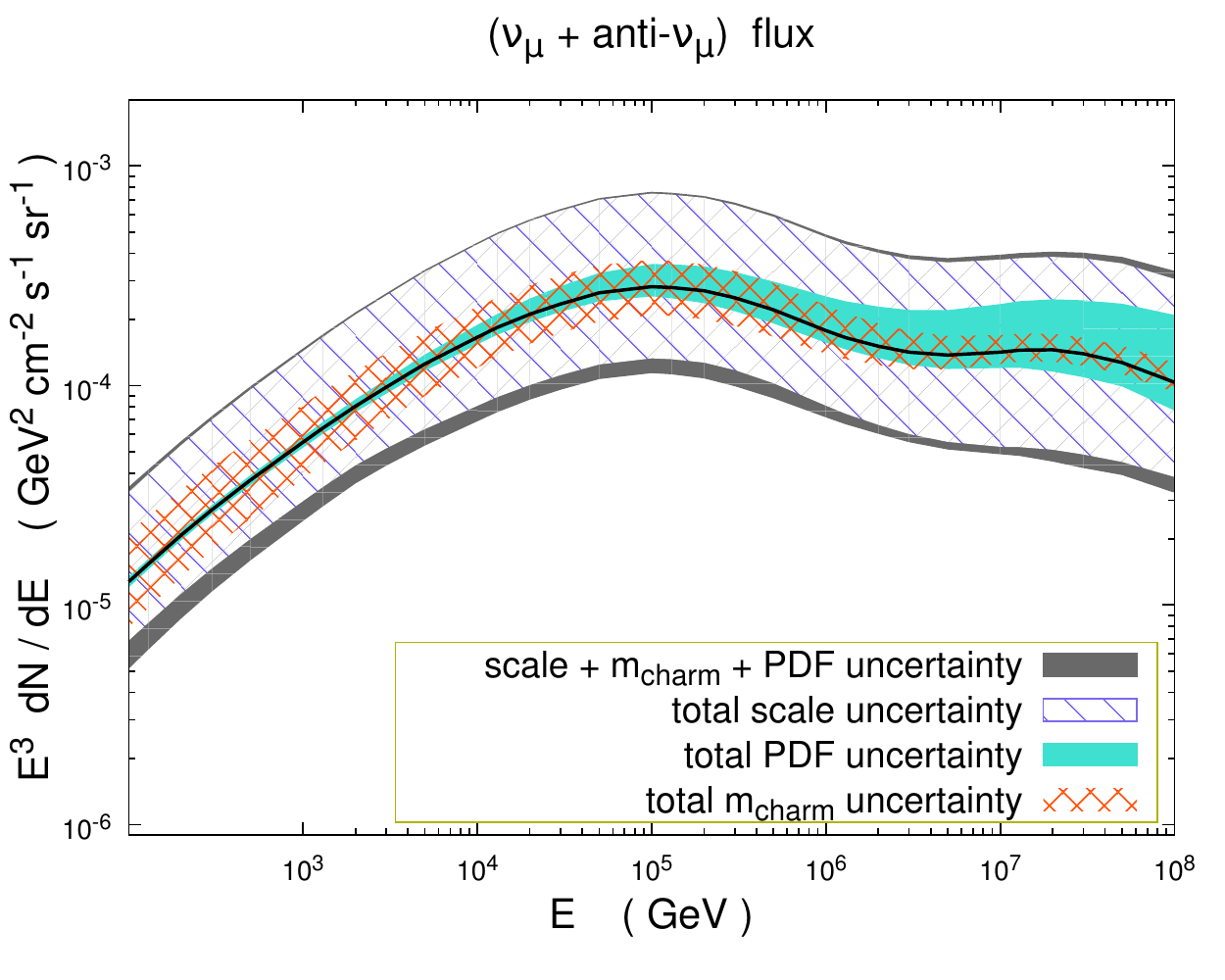}
\includegraphics[width=0.48\textwidth]{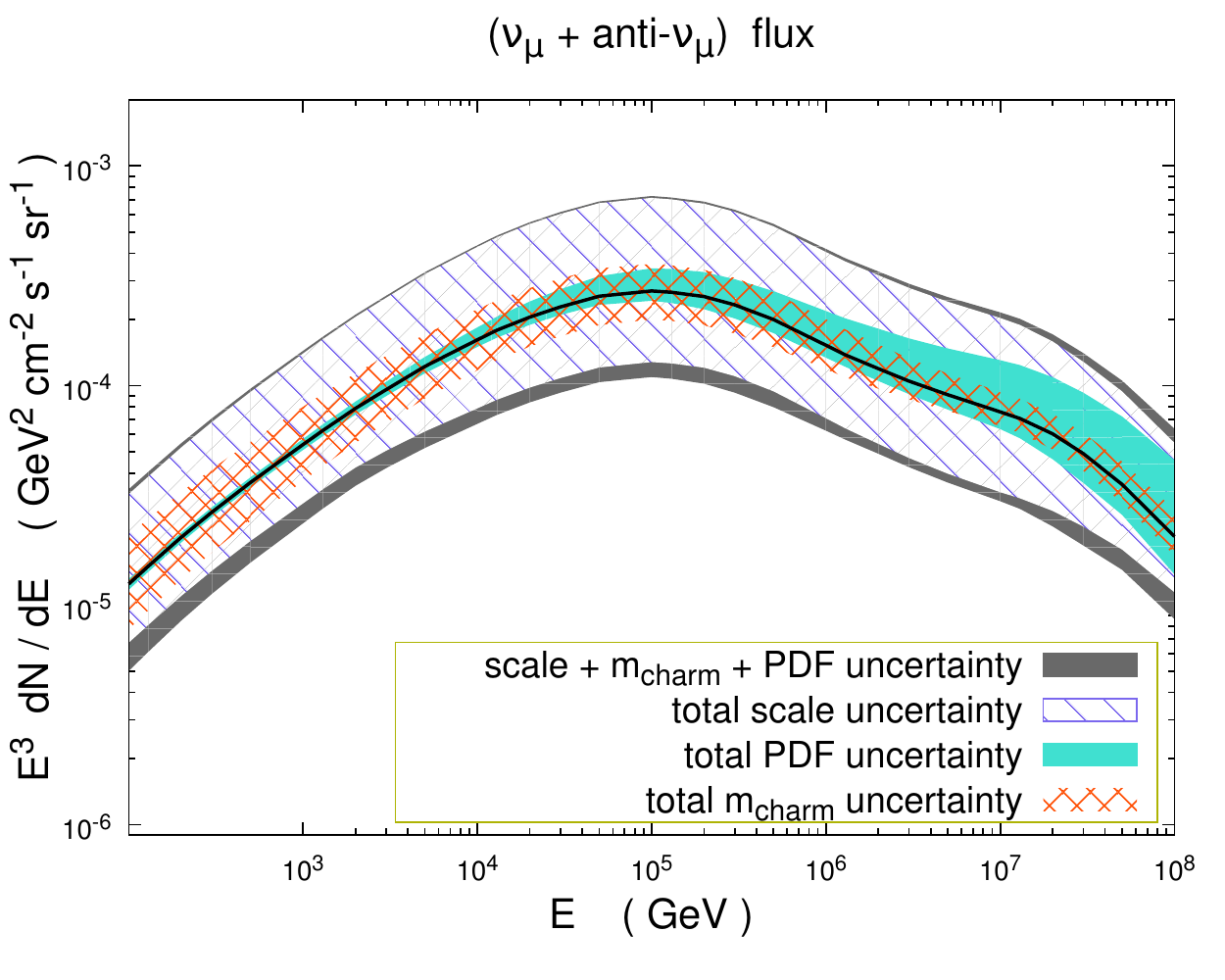}
\vspace{-0.5cm}
\end{center}
\caption{\label{prompt} Prompt ($\nu_\mu$~+~$\bar{\nu}_\mu$) spectra using as input the PROSA PDFs and two different popular choices for the composition of the CR primary spectrum (H3p and H3a all nucleon spectra). Various QCD uncertainties are shown.}
\end{figure}

The PROSA PDF fit has been applied in Neutrino Astronomy~\cite{Garzelli:2016xmx}.
In the following we focus on the computation of prompt neutrino fluxes, a relevant background for an accurate estimate of the astrophysical neutrino flux at Very Large Volume Neutrino Telescopes (VLV$\nu$T's). 
 The most energetic events recorded so far by VLV$\nu$T's have energies of a few PeV. Prompt neutrinos with these energies, are typically produced by the semileptonic decay of charmed mesons in $pp$ collisions at LHC energies. On the one hand, LHCb data allow to probe a limited $y$ range ($<$ 4.5), that allowed us to constrain the PDF ``only'' down to $x$ $\sim$ $10^{-6}$. On the other hand estimating the prompt neutrino fluxes requires a computation of open charm hadroproduction at even larger rapidities. Thus, making predictions for prompt neutrino fluxes at the PeV scale necessarily involves an extrapolation of the PDFs to $x$ values even lower than $10^{-6}$. 
However, at these energies, the relevant differential cross-section, $d \sigma/ d x_E$, where $x_E$ is the ratio of the energy of the produced $D$-meson with respect to the energy of the incident CR proton, is still dominated by collisions of partons with $x$ $>$ $10^{-6}$~\cite{Goncalves:2017lvq}. Thus, our predictions on prompt neutrino fluxes can be considered robust up to at least $E_{\nu}$ $\sim$ $\mathcal{O}$(PeV). For larger $E_{\nu}$, the uncertainties on prompt neutrino fluxes are increasingly driven by the reliability (or not) of the extrapolation of the PDFs to lower $x$ values. Additionally, further uncertainties, of astrophysical nature, related to our uncertain knowledge of the composition of the primary CR spectrum, become important for $E_{\nu}$~$\gtrsim$~$5~\cdot~10^{5}$~-~$10^{6}$~PeV.  
Warning the reader about these subtle aspects in the interpretation of the results, we provide in Fig.~\ref{prompt} our predictions for prompt ($\nu_\mu$~+~$\bar{\nu}_\mu$) fluxes, using the PROSA PDFs and two different hypotheses for the composition of the primary CR spectrum~\cite{Gaisser:2011cc, Gaisser:2013bla}. 
With the PROSA PDFs, differently from other PDF choices (see e.g. the discussion in Ref~\cite{Benzke:2017yjn}), the PDF uncertainties have become subleading with respect to the QCD uncertainties related to scale variation that dominate our prompt neutrino predictions.

\section{Acknowledgements}
I am grateful to S.~Alekhin, M. Benzke, T.~K.~Gaisser, B.~A.~Kniehl, G.~Kramer and F.~Riehn for interesting discussions which have contributed to shape part of this work and to the DESY for having partially supported my participation to the Workshop. 
\newpage

\bibliographystyle{polonica}
\bibliography{diffmv2}

\end{document}